\documentclass[aps, prc, reprint, superscriptaddress]{revtex4-1}
\usepackage{amsmath}
\usepackage{amsfonts}
\usepackage{amssymb}

\usepackage{graphicx}
\usepackage[caption=false]{subfig}

\usepackage[colorlinks]{hyperref}
\hypersetup{linkcolor=blue, citecolor=blue, filecolor=blue, urlcolor=blue}
\usepackage[all]{hypcap}

\usepackage[utf8]{inputenc}

\usepackage{romannum}
\usepackage{booktabs}
\RequirePackage{lineno}

\newcommand{\muB}{$\mu_{\rm B}$}
\newcommand{\sNN}{$\sqrt {{s_{\rm NN}}}~$}

\newcommand{\acceptane}{$0.4 < p_{\rm{T}} < 2.0$ GeV/$c$ }

\begin{document}
\title{Centrality selection effect on higher-order cumulants of net-proton multiplicity distributions in relativistic heavy-ion collisions}

\author{Arghya Chatterjee}
\author{Yu Zhang}
\author{Jingdong Zeng}
\affiliation{Key Laboratory of Quark \& Lepton Physics (MOE) and Institute of Particle Physics, Central China Normal University, Wuhan 430079, China}
\author{Nihar Ranjan Sahoo}
\affiliation{Shandong University, Qingdao, Shandong 266237, China}
\author{Xiaofeng Luo}
\email{xfluo@mail.ccnu.edu.cn}
\affiliation{Key Laboratory of Quark \& Lepton Physics (MOE) and Institute of Particle Physics, Central China Normal University, Wuhan 430079, China}


\begin{abstract}
We studied the centrality selection effect on cumulants (up to fourth order) and the cumulants ratios of net-proton multiplicity distributions in Au+Au collisions at $\sqrt{s_{\mathrm{NN}}}$ = 7.7, 19.6 and 200 GeV from UrQMD model. The net-proton cumulants are calculated with collision centralities by using charged particle multiplicity from different pesudorapidity ($\eta$) region. By comparing the results from various collision centralities, we found that the autocorrelation effects are not significant in the results with collision centralities "refmult-3" and "refmult-2", which are using mid-rapidity charged particles but excluding (anti-)protons and analysis region, respectively.  Furthermore, due to the contributions of spectator protons, we observed poor centrality resolution when using charged particles at forward $\eta$ region at low energies. This work can serve as a baseline for centrality selection of future fluctuations analysis in relativistic heavy-ion collisions. 
\end{abstract}
\maketitle
\section{Introduction}

One of the major goal of high-energy heavy-ion collision experiments is to explore the phase structure of the strongly interacting QCD matter~\cite{Aggarwal:2010cw}. The QCD phase structure can be represented as a function of temperature ($T$) and baryon chemical potential ($\mu_{B}$)~\cite{Rajagopal:1999cp}. QCD based model calculations predict that at large \muB~the transition from hadronic matter to Quark-Gluon Plasma (QGP) is of first order~\cite{Stephanov:2007fk,Bowman:2008kc}. The end point of the first order phase transition boundary is known as QCD critical point (CP), after which there is no genuine phase transition but a smooth crossover from hadronic to quark-gluon degrees of freedom~\cite{Aoki:2006we,Gupta:2011wh}. Many efforts has been made to find the signature and/or location of the CP, theoretically~\cite{Fodor:2004nz,deForcrand:2002hgr,Qin:2010nq,Xin:2014ela,Shi:2014zpa,Fischer:2014ata,Lu:2015naa,Zhang:2017icm,Bazavov:2017tot,Fu:2019hdw,Fischer:2018sdj,Li:2018ygx} and experimentally~\cite{Luo:2017faz}.  However, the location of the CP and even the existence of the CP have not been confirmed yet.
Experimental confirmation of the existence of the critical point will be a milestone for the study of QCD phase structure. 

One of the foremost method for the critical point search is through measuring the event-by-event higher-order fluctuations (called 'cumulants') of conserved quantities, such as net-charge ($Q$), net-baryon ($B$) and net-strangeness($S$), because of their divergence nature near the critical point~\cite{Stephanov:1998dy,Bzdak:2019pkr}. Due to the limitation of measuring neutral particles, experimentally we measured the cumulants of net-proton, and net-kaon as a proxy of net-baryon and net-strangeness respectively. The STAR experiment at RHIC, over past few years have measured the higher order cumulants up to forth order of net-proton~\cite{Aggarwal:2010wy, Adamczyk:2013dal, Luo:2015doi,Adam:2020unf}, net-charge~\cite{Adamczyk:2014fia} and net-kaon~\cite{Adamczyk:2017wsl} multiplicity distributions. Recently STAR has also reported the cross-cumulants between net-particles~\cite{Adam:2019xmk}. Theoretically $n$-th order cumulants are related to the $n$-th order thermodynamic susceptibilities as $C_{n} = VT^{3} \chi_{n}$, where $V$ and $T$ are the system volume and temperature, respectively. In order to compare the experimental measurements with theoretical susceptibilities, different cumulants ratios are constructed (like, $C_{2}/C_{1} = \chi_{2}/\chi_{1}, C_{3}/C_{2} = \chi_{2}/\chi_{1}$ and $C_{4}/C_{2} = \chi_{4}/\chi_{2}$, etc.). Cumulants values also related with the correlation length ($\xi$) of the matter created in the collisions as $\sigma^{2} = C_{2} = \xi^{2}; S = C_{3}/C_{2}^{3/2} = \xi^{4.5}; \kappa = C_{4}/C_{2}^{2} = \xi^{7}$~\cite{Stephanov:2008qz,Stephanov:2011pb}. One of the characteristic signatures of the QCD CP is the divergence of correlation length which gives a non-monotonic variation of these cumulant ratios as a function of \muB. The STAR experiment at RHIC has measured the cumulant ratios of net-proton, net-charge and net-kaon multiplicity distributions in Au+Au collisions at broad range of collision energies from 200 GeV down to 7.7 GeV, which correspond to a chemical freeze-out \muB~range from 20 to 420 MeV. Interestingly, the forth-order net-proton cumulant ratio ($\kappa\sigma^{2}=C_{4}/C_{2}$) for most central 0-5\% collisions shows a non-monotonic variation as a function of collision energy~\cite{Luo:2015doi}. 

To understand the underlying physics associated with this measurement, we need to perform careful studies on the background contributions, such as the effects from initial volume fluctuations, the detector efficiency and the effects of centrality selection. Some of the effects are discussed before~\cite{Luo:2013bmi, Chatterjee:2016mve, Zhang:2019lqz, Westfall:2014fwa,Zhou:2018fxx,Jia:2020tvb}. Collision centralities can be quantified by impact parameter ($b$) or number of participant nucleons ($N_{part}$). Unfortunately, in experiment we cannot directly measure such geometrical variables. As the particle multiplicities depend on initial geometry, so the collision centrality in heavy-ion collisions is usually determined by the charge particle multiplicities. The centrality resolution is determined by the multiplicities and kinematics of the charged particles used in the centrality definition. 
As the bad centrality resolution will introduce larger volume fluctuations and enhance the higher order cumulants, a good centrality resolution of charged particle centrality definition is very important for fluctuation analysis. 
On the other hand, there is so called autocorrelation effect~\cite{Luo:2013bmi}, which indicates that values of higher order cumulants will be suppressed if the charged particles involved in centrality definition are also used in the cumulant calculations. To avoid the autocorrelation, particles from different kinematic region are proposed to define the collision centralities. Experimentally,  a dedicate Event Plane Detector (EPD)~\cite{Adams:2019fpo} has been built and installed in the froward region ($2.1<|\eta|<5.1$) of the STAR experiment. The EPD will be used for event plane and centrality determination in the second phase of the Beam Energy Scan program (BES-II,2019-2021) at RHIC. It has been proposed that the centrality selection by using Event Plane Detector (EPD) will strongly suppress the effect of autocorrelation in the fluctuation analysis~\cite{Shanmuganathan:2017nmb}. In this work, we will demonstrate the variation of net-proton cumulants values by selecting centralities from different central as well as forward region using UrQMD model.

The paper is organized as follows. In section II, we briefly discuss the UrQMD model used for this analysis. In section III, we introduce the observables presented here. The centrality selection is discussed in section IV. In section V, we present cumulants ($C_{1}$-$C_{4}$) of net-proton multiplicity distributions for different centrality definition in Au+Au collisions at \sNN = 7.7, 19.6 and 200 GeV using UrQMD model. Finally in section VI, we present a summary of this work.

\section{The UrQMD Model}

The Ultra Relativistic Quantum Molecular Dynamic (UrQMD) is a microscopic transport model~\cite{Bass:1998ca, Bleicher:1999xi}. In this model, the space-time evolution of the fireball is studied in terms of excitation of color strings which fragment further into hadrons, the covariant propagation of hadrons and resonances which undergo scatterings and finally the decay of all the resonances. UrQMD model has been quit successful and widely applied towards heavy-ion phenomenology~\cite{Bleicher:1999xi, Bleicher:1998wu}. Previously, this model has been used to compute several cumulants and studied different effects of experimental limitations~\cite{Bleicher:2000ek, Haussler:2005ei, Sahoo:2012wn, Luo:2013bmi, Chatterjee:2016mve, Xu:2016qjd,He:2017zpg, Mukherjee:2017elm, Zhou:2017jfk}. The choice of acceptance window plays an important role to such studies. The initial distributions of net-baryon ($N_{B}$) in rapidity is a consequence of the baryon stopping phenomenon which strongly depends on collision energy. As a result, the mid-rapidity region for high \sNN is free of $N_{B}$ while, at lower \sNN, most of the $N_{B}$ are deposited in mid-rapidity region. This collision energy dependence baryon stopping phenomenon is dynamically included in the UrQMD model. More details about the UrQMD model can be found in the reference~\cite{Bass:1998ca, Bleicher:1999xi}. In this study, we have used six million events per beam energy for Au+Au collisions at \sNN = 7.7, 19.6 and 200 GeV. Using this simulated events, we measure the cumulants of event-by-event net-proton ($N_{p-\bar{p}}$) multiplicity distributions within the kinematic acceptance $|y|<0.5$ and \acceptane. The same kinematic acceptance have been used in the net-proton cumulant analysis by STAR experiment~\cite{Luo:2015doi}.

\section{Observables}

In statistics, any distribution can be characterized by different order moments or cumulants ($C_{n}$) and can be expressed via generating function~\cite{Kitazawa:2017ljq} as,

\begin{eqnarray}
C_{n} = \frac{\partial^{n}}{\partial \alpha^{n}} K(\alpha)|_{\alpha = 0},
\end{eqnarray}

where $K(\alpha) = \ln(M(\alpha))$ and $M(\alpha) =  \langle e^{\alpha N} \rangle$ are the cumulant and moment generating functions, respectively. $N$ is the event-by-event net-quantity (here net-proton number, $N_{p} = N_{p} - N_{\bar{p}}$) and $\langle ... \rangle$ represents an average over events. Then the various order cumulants can be expressed as, 

\begin{eqnarray}
C_{1} &=& \langle N \rangle, \\
C_{2} &=& \langle (\delta N)^{2} \rangle, \\
C_{3} &=& \langle (\delta N)^{3} \rangle, \\
C_{4} &=& \langle (\delta N)^{4} \rangle - 3\langle (\delta N)^{2} \rangle,
\end{eqnarray}
where $\delta N = N - \langle N \rangle$ represents the deviation of $N$ from its average value.  $\langle (\delta N)^{m} \rangle$ is also called the $m$-th order central moment. Thermodynamically the cumulants are connected to the corresponding susceptibilities by 

\begin{eqnarray}
 C_{n} = \frac{\partial^{n} \ln(Z(V,T,\mu))}{\partial \mu^{n}} = VT^{3} \chi_{n},
\end{eqnarray}

The cumulant ratios between different orders can be constructed to cancel the volume term. This cumulants ratios are measured experimentally and compared to the susceptibility ratios~\cite{Luo:2017faz, Cheng:2008zh},

\begin{eqnarray}
\frac{\sigma^{2}}{M} = \frac{C_{2}}{C_{1}} = \frac{\chi_{2}}{\chi_{1}},~S\sigma = \frac{C_{3}}{C_{2}} = \frac{\chi_{3}}{\chi_{2}},~\kappa\sigma^{2} = \frac{C_{4}}{C_{2}} = \frac{\chi_{4}}{\chi_{2}},   
\end{eqnarray}

With above definitions, we have studied various cumulants (up to forth order) and cumulant ratios of event-by-event net-proton multiplicity distributions from UrQMD model with different centrality selection. 
\begin{figure*}[htp!]
	\centering 
	\includegraphics[width=0.95\textwidth]{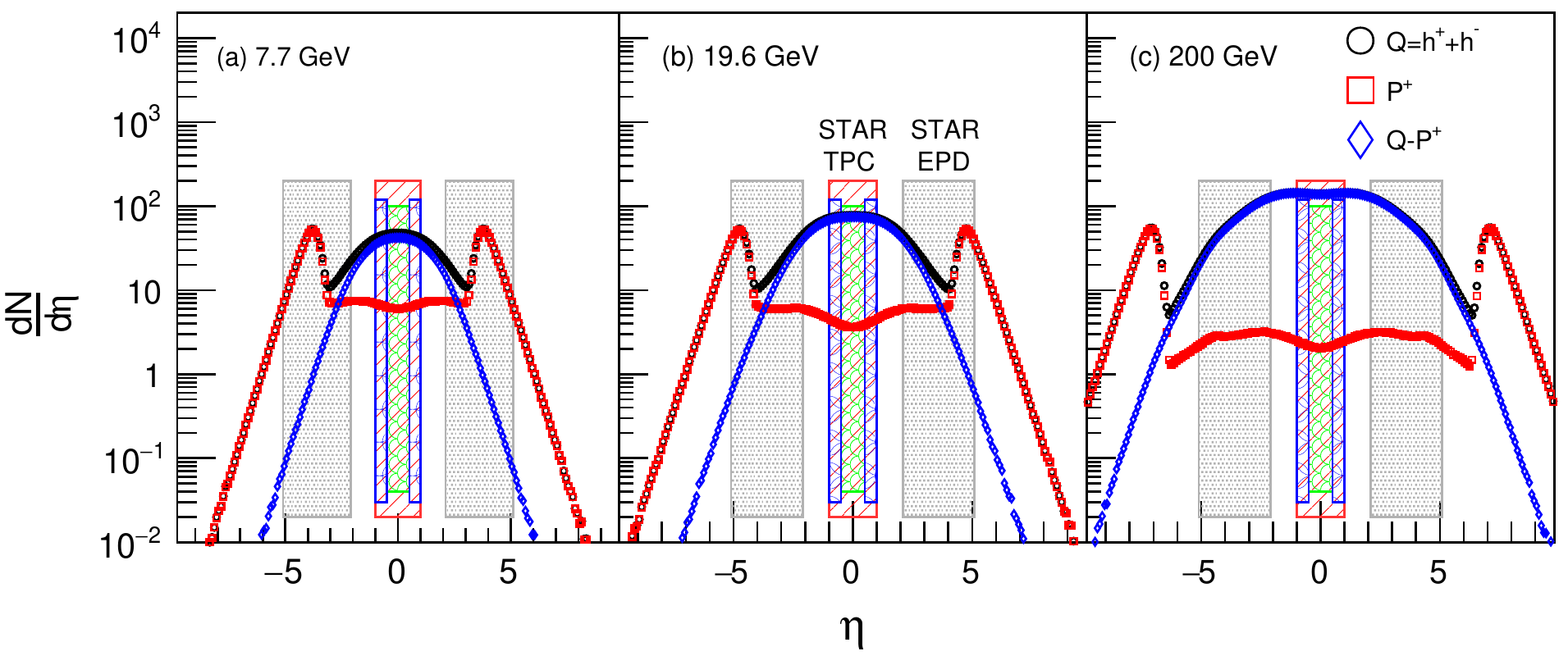}
	\caption{(Color online) The $dN/d\eta$ distributions for charged particle, proton and (number of charged particles - protons number) multiplicity in minimum bias Au+Au collisions at \sNN= 7.7, 19.6 and 200 GeV. The bands at different color correspond to different pesudorapidity ($\eta$) regions for centrality selection.}
	\label{pseudorapidity}
\end{figure*}

In heavy-ion collisions, we cannot directly measure the geometrical variables, such as impact parameter.  The collision centrality in heavy-ion collisions is usually determined through charged particle multiplicities, in which the smallest centrality bin is a single multiplicity value. However, for better statistical significance, we report the cumulant results for a wider centrality bins, like 0-5\% (most central) or 70-80\% (peripheral). But, this particle multiplicities not only reflects the initial geometry but also depends on different physics process. This also correspond that the measured observables $N_{ch}$ and geometrical variable ($b$) is not one-to-one correspondence. A fixed $N_{ch}$ may come from different initial geometry. This variation even become large when we use wider 5\% or 10\% centrality class. To reduced the variation for wider centrality bins, a so called Centrality Bin Width Correction (CBWC) technique is applied in cumulant analysis~\cite{Luo:2013bmi}. The techniques for this corrections are follows. We first calculate different cumulants ($C_{n}$) in each bin of unit multiplicity and then weight the cumulants by the number of events in each bin over a desired centrality class. The method can be expressed as,

\begin{eqnarray}
C_{n} = \frac{\sum_{i}n_{i}C_{n_{i}}}{\sum_{i}n_{i}} = \sum_{i} \omega_{i}C_{n_{i}},
\label{cbwc}
\end{eqnarray}
where $C_{n_{i}}$ is the cumulant value measured in the $i^{th}$ multiplicity bin. $n_{i}$ and $\omega_{i} (= n_{i}/\sum_{i}n_{i})$ are the number of events and the weight factor for $i^{th}$ multiplicity bin. It was shown that the CBWC can effectively suppress the volume fluctuations within a wide centrality bin~\cite{Luo:2013bmi}. However, even CBWC is applied,  there could be still residual volume fluctuations if the centrality resolution is not good. 
Another centrality selection related artifact is so called autocorrelation effect, which is due to the correlations between particles used in the centrality selection and the cumulant calculations. For example, a typical autocorrelation effect is caused by the fact that some of the particles involved in the cumulant analysis are also used for the centrality selection. In the STAR experiment, to avoid the autocorrelation effect in net-proton and net-charge fluctuation measurements, collision centralities are carefully selected, the so called "refmult-3" and "refmult-2", respectively. 
In refmult-3 definition, the collision centrality is determined by the measured charged particle multiplicities ($N_{ch}$) within $|\eta|<1.0$ excluding protons and antiprotons, while the refmult-2 centrality is defined by the measured charged particles within $0.5<|\eta|<1.0$. In BES-II, a dedicate forward Event Plane Detector (EPD) with coverage $2.1<|\eta|<5.1$ will be installed at the STAR experiment. The EPD will collect the ionization signals of charged particles and 
allows us to define the collision centrality in the forward rapidity (like EPD region in STAR) instead of mid-rapidity region~\cite{Shanmuganathan:2017nmb}. We will demonstrate the effects of autocorrelations by selecting centralities from different region and discuss the effects of the spectator protons in the centrality selection with forward charged particles.

Figure~\ref{pseudorapidity} shows the pseudorapidity ($\eta$) distributions ($dN/d\eta$) of charged particle, proton ($p$) and $N_{ch} - p$ multiplicities for the min-bias Au+Au collisions at \sNN = 7.7, 19.6 and 200 GeV. The bands at different $\eta$ regions correspond to the acceptance of STAR Time Projection Chamber (TPC) and EPD, respectively.  The particle pseudorapidity ($\eta$) distribution is not uniform through out the acceptance for all energies. At the forward pseudo-rapidity region from 2 to 5 unit, one can find lots of spectator contributions at low energies. We can see two peaks structures around $|\eta|$ = 7-8 for 200 GeV, which correspond to the spectator protons. As we go towards low beam energies, the peaks shifted towards central $\eta$ region, like for 7.7 GeV the peak is around $|\eta|$ = 3.5. For the $\eta$ window from 2.1 to 5.1 unit at 200 GeV,  the charged particles are mostly contributed from the produced particles. However, if we go towards lower beam energies, charged particles in that range are dominated by spectator protons, as most of the spectator protons are around beam rapidity, 

\begin{figure}[htp!]
	\centering 
\hspace{-1.2cm} 
	\includegraphics[width=0.52\textwidth]{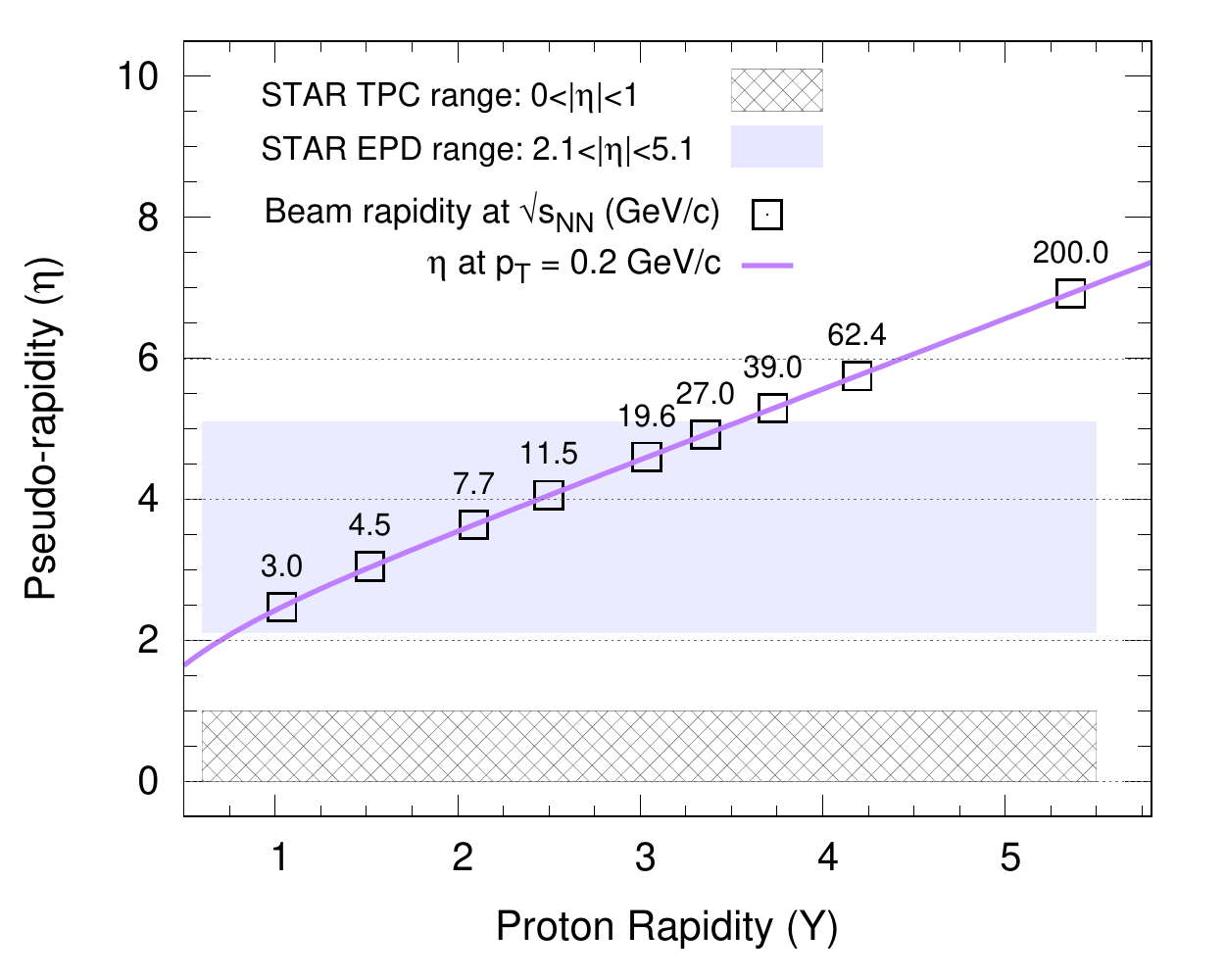}
	\caption{(Color online) Beam rapidity values as a function of center of mass energy. Central and forward detector region for centrality selection using charged-particle multiplicity represented in band. The relation of $p_{T}$, pseudo-rapidity ($\eta$) and rapidity ($y$) is ${p_T} = m_{0}/\sqrt {{{\sinh }^2}\eta /{{\sinh }^2}y - 1}$, where the $m_{0}$ is the particle rest mass.}
	\label{beamrapidity}
\end{figure}

Figure~\ref{beamrapidity} shows the beam rapidity for different center of mass energies and the corresponding $\eta$ values at $p_{T}=0.2$ GeV/c. We found that at low energies between 3 and 27 GeV, the protons with beam rapidity and $p_{T}=0.2$ GeV/c will fall into the $\eta$ coverage of the STAR EPD ($2.1<|\eta|<5.1$), which will also leave ionization signals in the EPD as other produced charge particles. Due to lack of particle identification capability of EPD, we have difficulties to isolate the signals of spectator protons, especially at low energies. The contamination of the spectator protons will distort the correlations between the charged particle signals and the collision centrality. Consequently, the collision centrality determined from EPD will have poor resolution, which will enhance the volume fluctuations within a centrality bin. In the following, we will demonstrate the effect of the spectator protons on the centrality selection and net-proton cumulant analysis.  
\section{Centrality selection}
\begin{table*}[htp!]
	\begin{center}
		\begin{tabular}{ c c}
			\hline
			Identify & Definition \\
			\hline
			centrality-b & impact parameter \\
			refmult-1 & $N_{ch}$ within $|\eta|<0.5$  \\
			refmult-2 & $N_{ch}$ within $0.5<|\eta|<1.0$  \\
			refmult-3 & $N_{ch}-p$ within $|\eta|<1.0$  \\
			Fwd-All & $N_{ch}$ within $2.1<|\eta|<5.1$  \\
			Fwd-1 & $N_{ch}$ within $2.1<|\eta|<3.0$  \\
			Fwd-2 & $N_{ch}$ within $3.0<|\eta|<4.0$  \\
			Fwd-3 & $N_{ch}$ within $4.0<|\eta|<5.0$  \\
			Fwd-All $-$ p & $N_{ch}-p$ within $2.1<|\eta|<5.1$  \\
			\hline
		\end{tabular}
		\caption {Definition of different centrality selection methods used in this work.}
		\label{table:centrality}
	\end{center}
\end{table*}

\begin{figure*}[htp!]
	\centering 
	\includegraphics[width=0.9\textwidth]{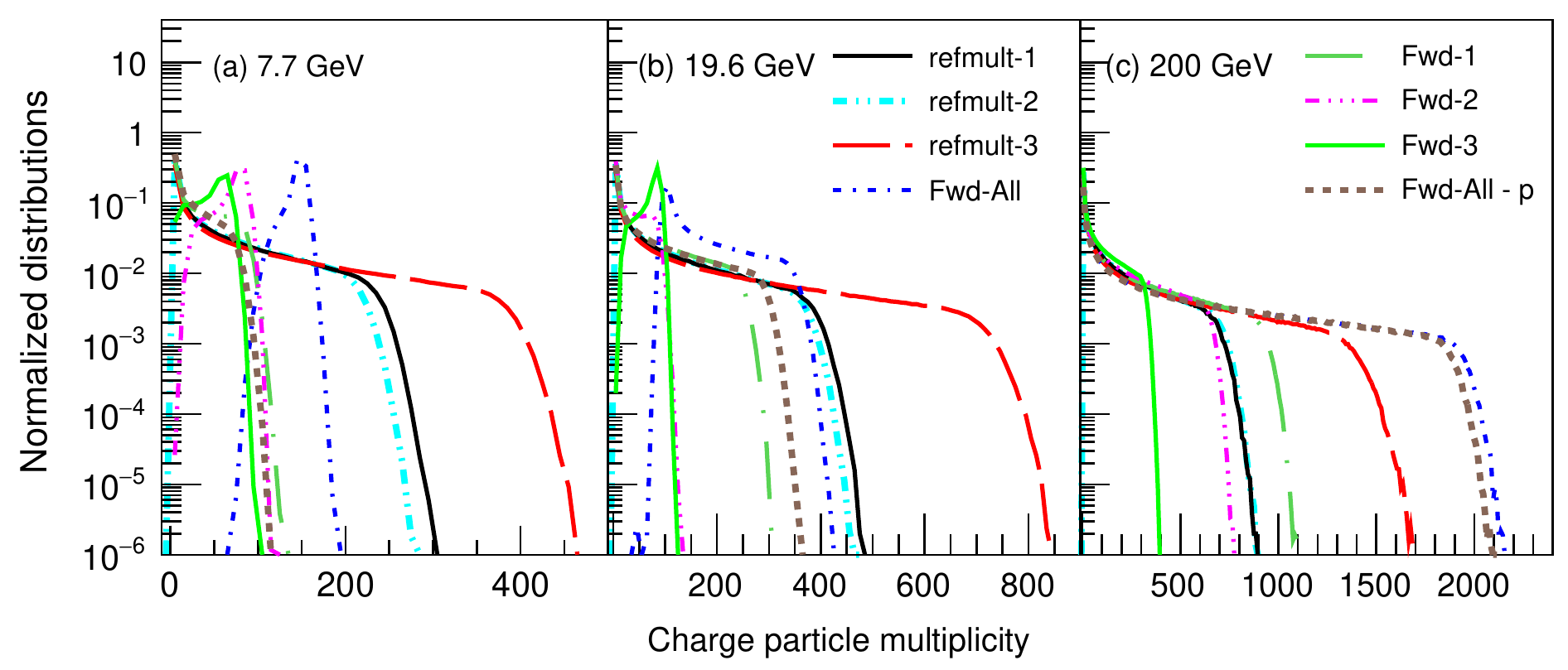}
	\caption{(Color online) Normalized distributions for charged particle multiplicities in different $\eta$-window in Au+Au collisions at \sNN = 7.7, 19.6 and 200 from UrQMD model.}
	\label{multiplicity}
\end{figure*}

In this work we select centralities using charge particle multiplicities from different $\eta$ region. The definitions of different centrality selections are listed in Table~\ref{table:centrality}. We further subdivide the forward region "Fwd-All" range in three region : (a) Fwd-1, pseudorapidity acceptance $2.1<|\eta|<3$, (b) Fwd-2 within $3<|\eta|<4$, and (c) Fwd-3 within $4<|\eta|<5$. Figure~\ref{multiplicity} shows the minimum bias charged particle multiplicity distributions at \sNN = 7.7, 19.6 and 200 GeV from different acceptance regions as listed in Table~\ref{table:centrality}. We select 9 different centrality classes: 0-5\% (top central), 5-10\%, 10-20\%, 20-30\%, 30-40\%, 40-50\%, 50-60\%, 60-70\% and 70-80\% from the area percentile of the multiplicity distributions. We can find that at \sNN = 7.7 and 19.6 GeV the multiplicity distributions at forward region behaves differently than central region. 
This is mainly caused by the spectator protons, which are positively correlated with impact parameter and are opposite to the trend of the produced charged particle multiplicity distributions. 
As shown in the Fig.~\ref{multiplicity}, if we exclude the protons from forward region, then the trend of the distributions looks like a hose-tail shaped similar to the central ones. 

\begin{figure*}[htp!]
	\centering 
	\includegraphics[width=0.95\textwidth]{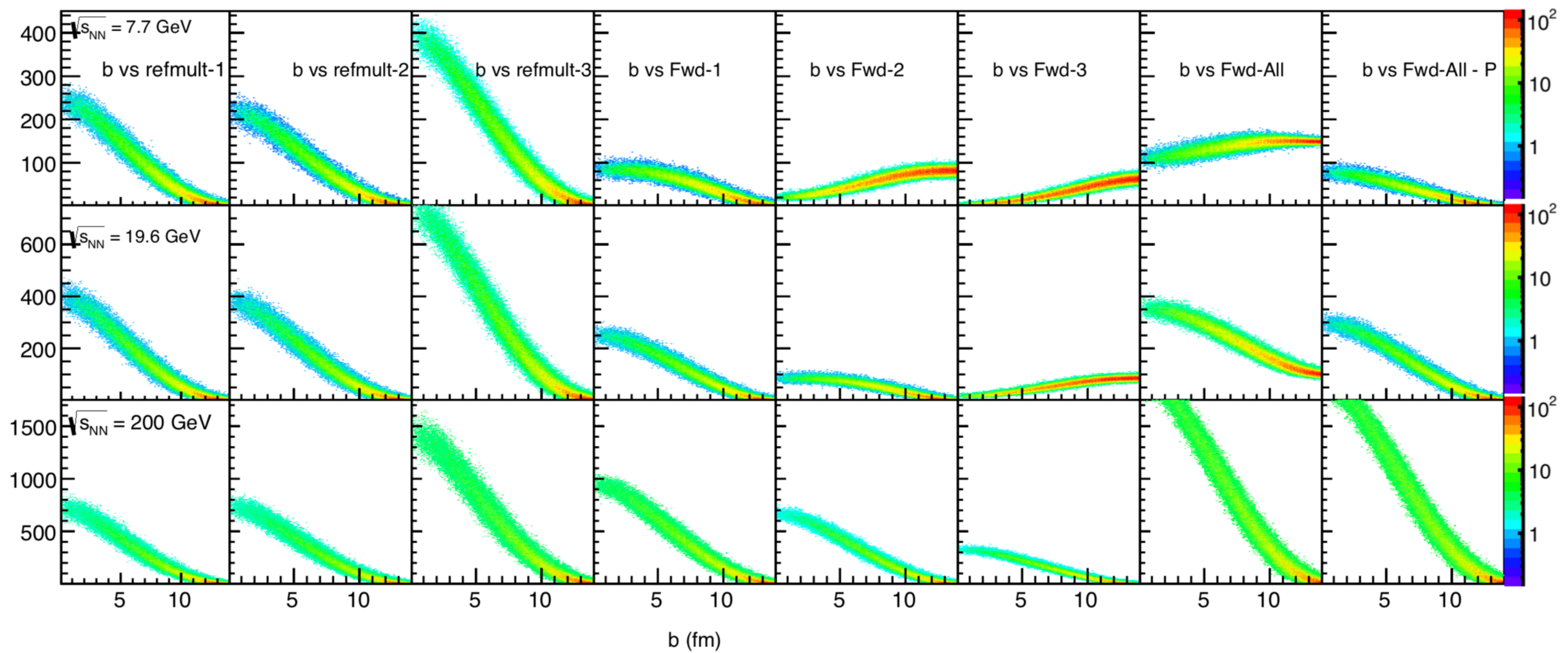}
	\caption{(Color online) Correlations between multiplicities in different $\eta$ window used for the centrality definitions and impact parameter in Au+Au collisions at \sNN= 7.7, 19.6 and 200 GeV from UrQMD model.}
	\label{correlation}
\end{figure*}

\begin{figure*}[htp!]
	\centering 
	\includegraphics[width=0.85\textwidth]{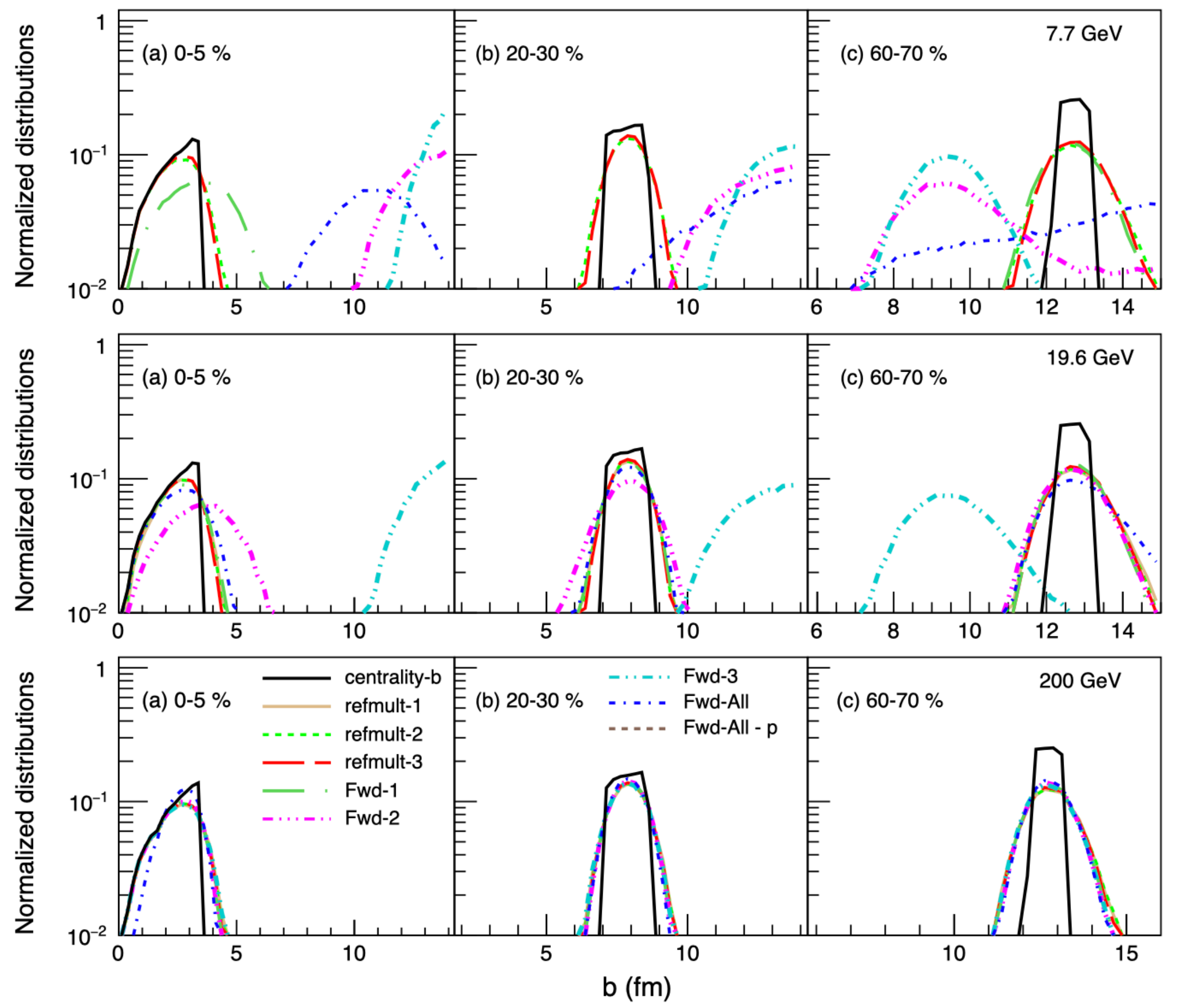}
	\caption{(Color online) The impact parameter (b) distributions in different centrality definitions for three different centrality classes ((a) 0-5\%, (b) 20-30\% and (c) 60-70\%) at $\sqrt{s_{\mathrm{NN}}}$ = 7.7, 19.6 and 200 GeV.}
	\label{impacdis}
\end{figure*}

\begin{figure*}[htp!]
	\centering 
	\vspace{0.5cm} 
	\includegraphics[width=0.9\textwidth]{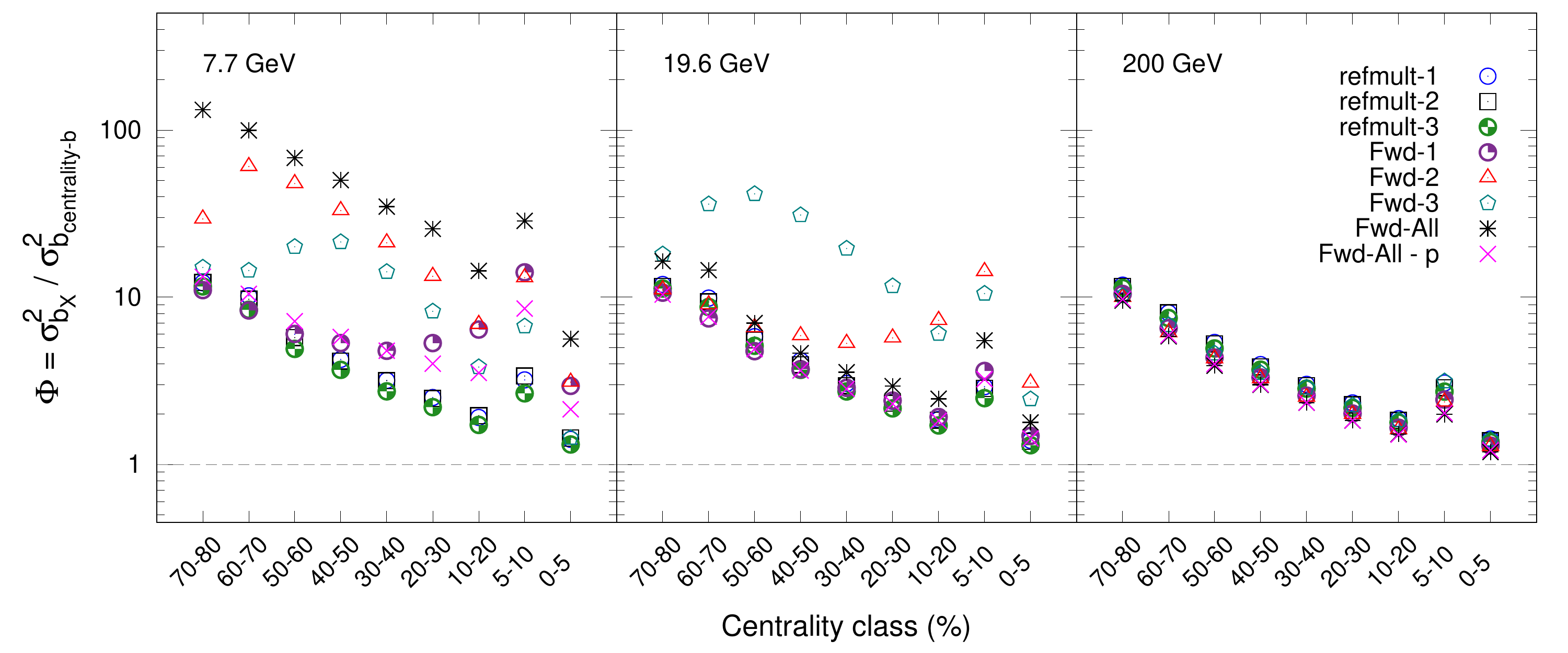}
	\caption{(Color online) The centrality dependence of the $\Phi (= \sigma^{2}_{b_{\text{X}}} / \sigma^{2}_{b_{\text{centrality-b}}})$ of impact parameter distributions for Au+Au collisions at \sNN = 7.7, 19.6 and 200 GeV in UrQMD model with different centrality definitions.}
	\label{resolution}
\end{figure*}

Figure~\ref{correlation} shows the two dimension correlation plots between the charged particle multiplicity distributions in different acceptance and the impact parameter. It was found that at lower energies the multiplicities within $2.1<|\eta|<5.1$ is positively correlated with impact parameter, i.e., we get more particles in peripheral collisions than central collisions, which is mainly contributed from the spectator protons. Figure~\ref{impacdis} shows the impact parameter distributions for three different centrality classes in Au+Au collisions at $\sqrt{s_{NN}}$ = 7.7, 19.6 and 200 GeV using different centrality definitions. To compare the centrality resolution between different centrality definitions, we define a quantity $\Phi (= \sigma^{2}_{{b_\text{X}}} / \sigma^{2}_{{b_\text{centrality-b}}})$ as shown in Fig.~\ref{resolution}, where the "\text{X}" is different centrality definitions using $N_{ch}$ as discussed before. Here, $\sigma^{2}_{b_{\text{X}}}$ is the variance in impact parameter distribution in a centrality class from "X" centrality definition whereas the $\sigma^{2}_{b_{\text{centrality-b}}}$ represents the variance in impact parameter distribution with the centrality defined by the impact parameter ($b$) itself. So larger values in $\Phi$ corresponds to a poorer resolution than smaller $\Phi$ values.  

\begin{figure*}[htp!]
	\centering 
	\includegraphics[width=0.9\textwidth]{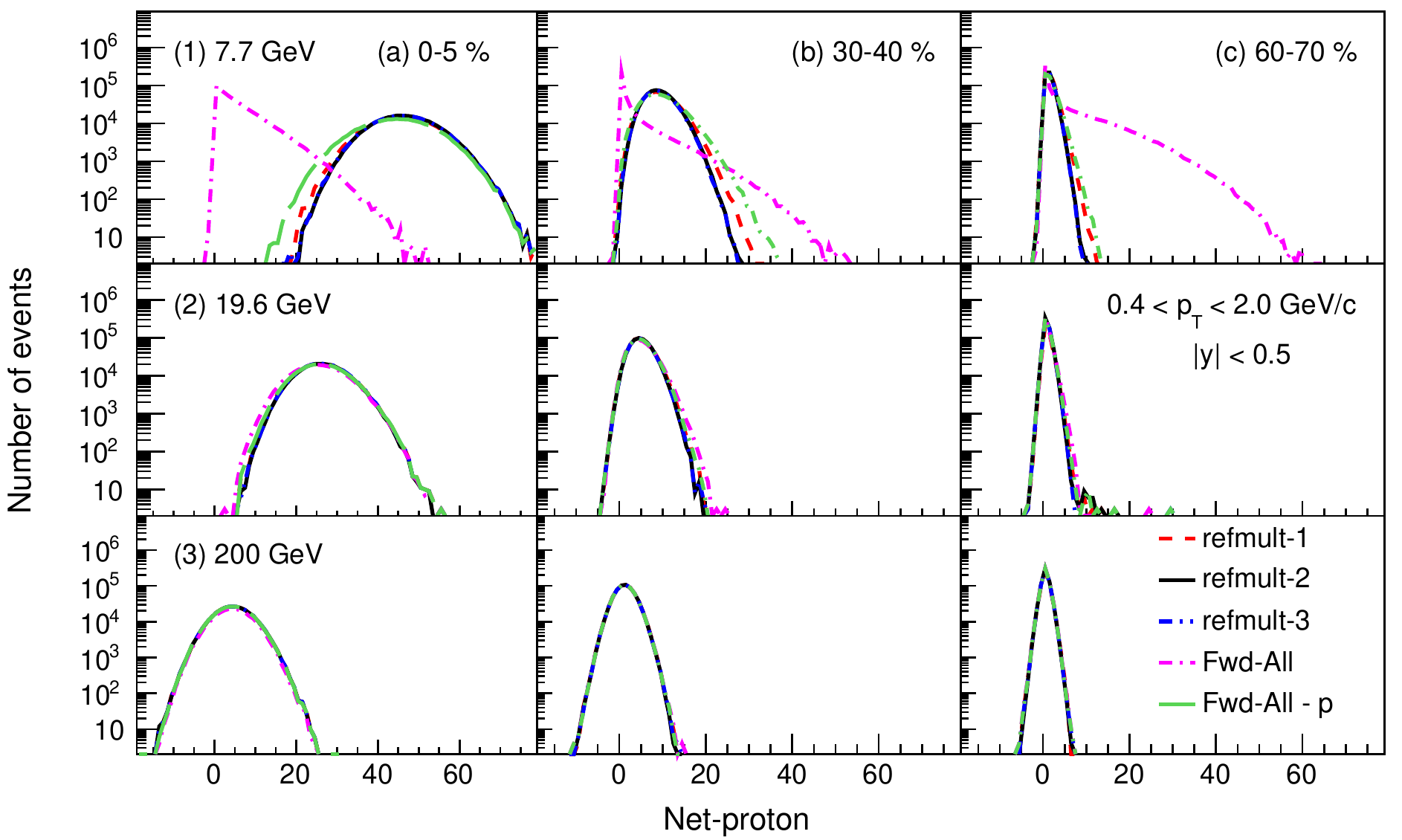}
	\caption{(Color online) Event-by-event distributions of net-proton multiplicity distribution for Au+Au collisions at \sNN= 7.7, 19.6 and 200 GeV for different centrality selection methods.}
	\label{netplot}
\end{figure*}

We find that the resolution in "refmult-3" definition is better for all energies followed by "refmult-2" and "refmult-1". At \sNN = 7.7 and 19.6 GeV, the resolution becomes poorer as we go towards larger $\eta$ region. This is due to the spectator contributions. It is also observed that the resolution get improved if we select centrality from forward region by excluding protons. But still $\Phi$ value is large at 7.7 GeV because of smaller number of produced particles in that region. We can also observed that the centrality resolutions are always better in central collisions than those from peripheral collisions.

\section{Results}
In this study, we compare the cumulants and their ratios of event-by-event net-proton multiplicity distributions within the kinematic acceptance $|y|<0.5$ and \acceptane for different centrality definitions as discussed in the previous section. Figure~\ref{netplot} shows the event-by-event net-proton multiplicity distributions for Au+Au collisions at \sNN = 7.7, 19.6 and 200 GeV for three centralities (0-5\%, 30-40\% and 60-70\%). We can find that the mean and width are larger for central than peripheral collisions. The mean values of net-proton distributions shifted towards zero as the energy increases. At 200 GeV, the net-proton distributions from all the centrality sets are very similar. At 7.7 GeV, the net-proton distributions for "Fwd-All" centrality case looks completely different. This is mainly caused by the distortion of the spectator protons in the centrality definition "Fwd-All". As shown in Fig.~\ref{impacdis}, due to the positive correlations between the number of spectator protons and the impact parameter, the impact parameter distribution from "Fwd-All" contains more peripheral events (large b values) in 0-5\% centrality class than that of 60-70\% centrality class. One needs to keep in mind that the raw net-proton distributions shown in Fig.~\ref{netplot} are not directly used to calculate various order cumulants and needs to apply the CBWC to suppress volume fluctuations in a wide centrality bin. 

\begin{figure*}[htp!]
	\centering 
	\includegraphics[width=0.9\textwidth]{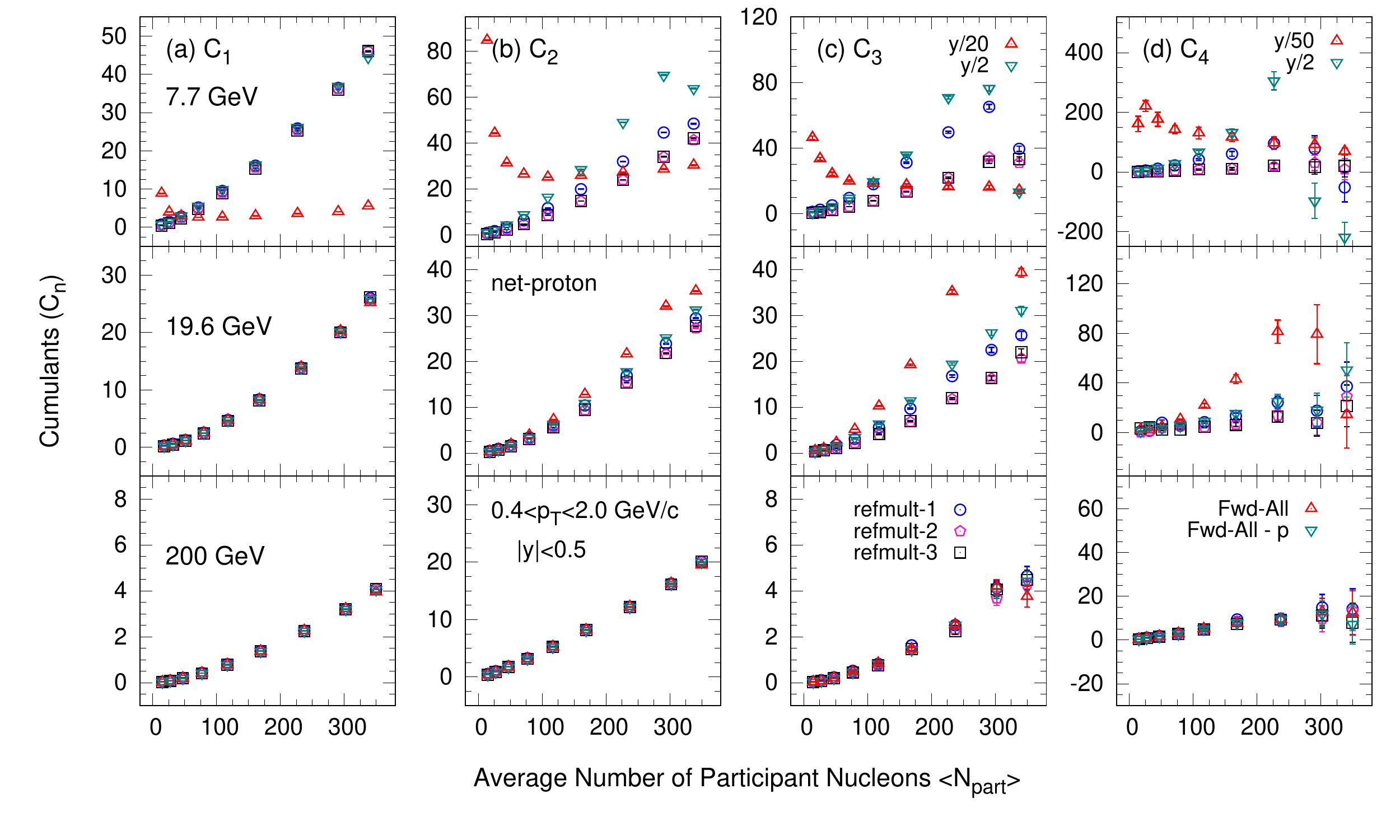}
	\caption{(Color online) Centrality dependence of cumulants ($C_{1} \sim C_{4}$) of net-proton multiplicity distributions within $|y|<0.5$ and $0.4<p_{T}<2.0$ GeV/$c$ for Au+Au collisions at \sNN= 7.7, 19.6 and 200 GeV for different centrality selection methods. The third and fourth order cumulant from "Fwd-All" and "Fwd-All-p" centrality definitions at  \sNN=7.7 GeV are scaled with different factors to compare with other cases. }
	\label{cumulants}
\end{figure*}

\begin{figure*}[htp!]
	\centering 
	\includegraphics[width=0.9\textwidth]{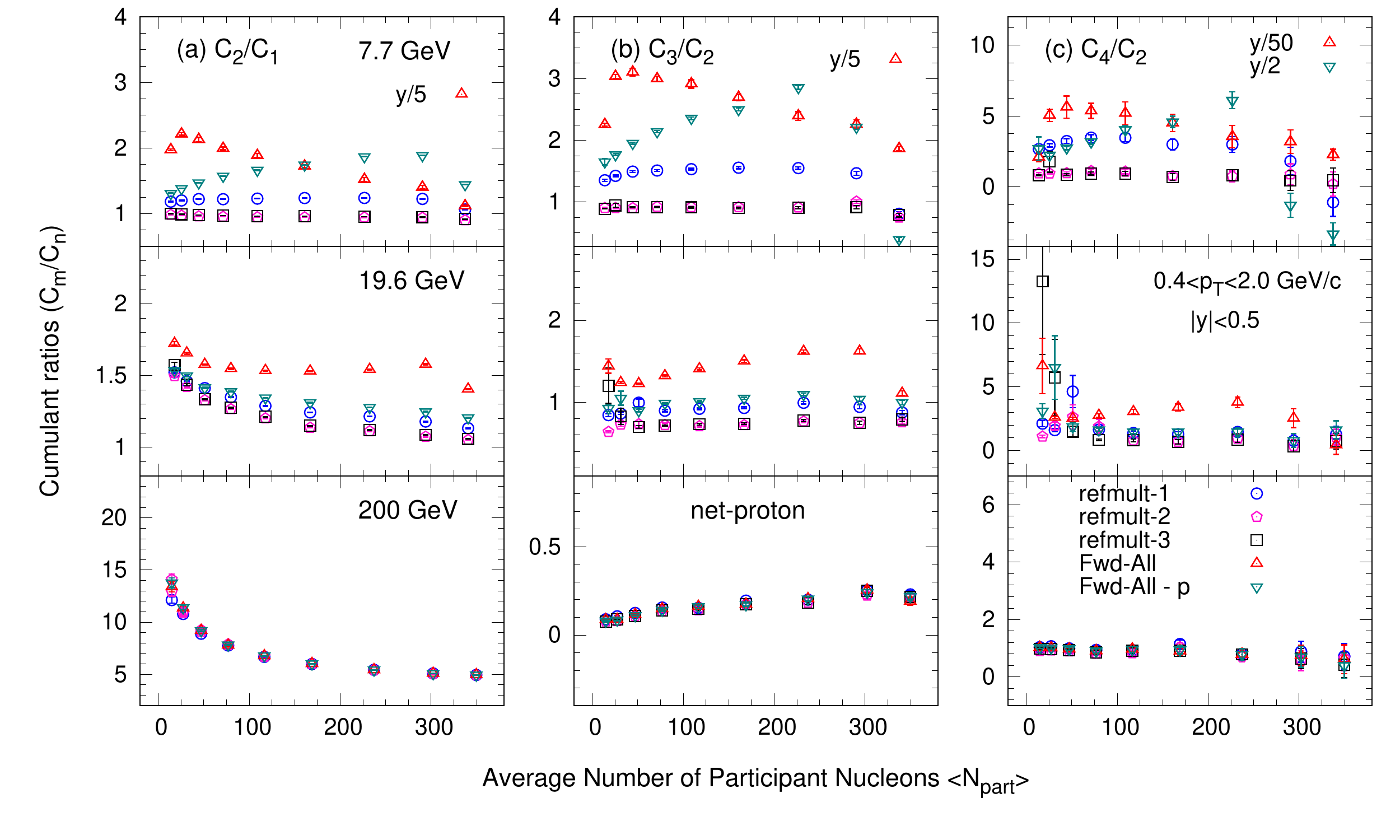}
	\caption{(Color online) Centrality dependence of cumulant ratios ($C_{2}/C_{1}$, $C_{3}/C_{2}$ and $C_{4}/C_{2}$) of net-proton multiplicity distributions within $|y|<0.5$ and $0.4<p_{T}<2.0$ GeV/$c$ for Au+Au collisions at \sNN= 7.7, 19.6 and 200 GeV for different centrality selection methods.  The second, third and fourth order cumulant from "Fwd-All" and "Fwd-All-p" centrality definitions at  \sNN=7.7 GeV are scaled with different factors to compare with other cases.}
	\label{ratio}
\end{figure*}

\begin{figure}[htp!]
	\centering 
	\includegraphics[width=0.5\textwidth]{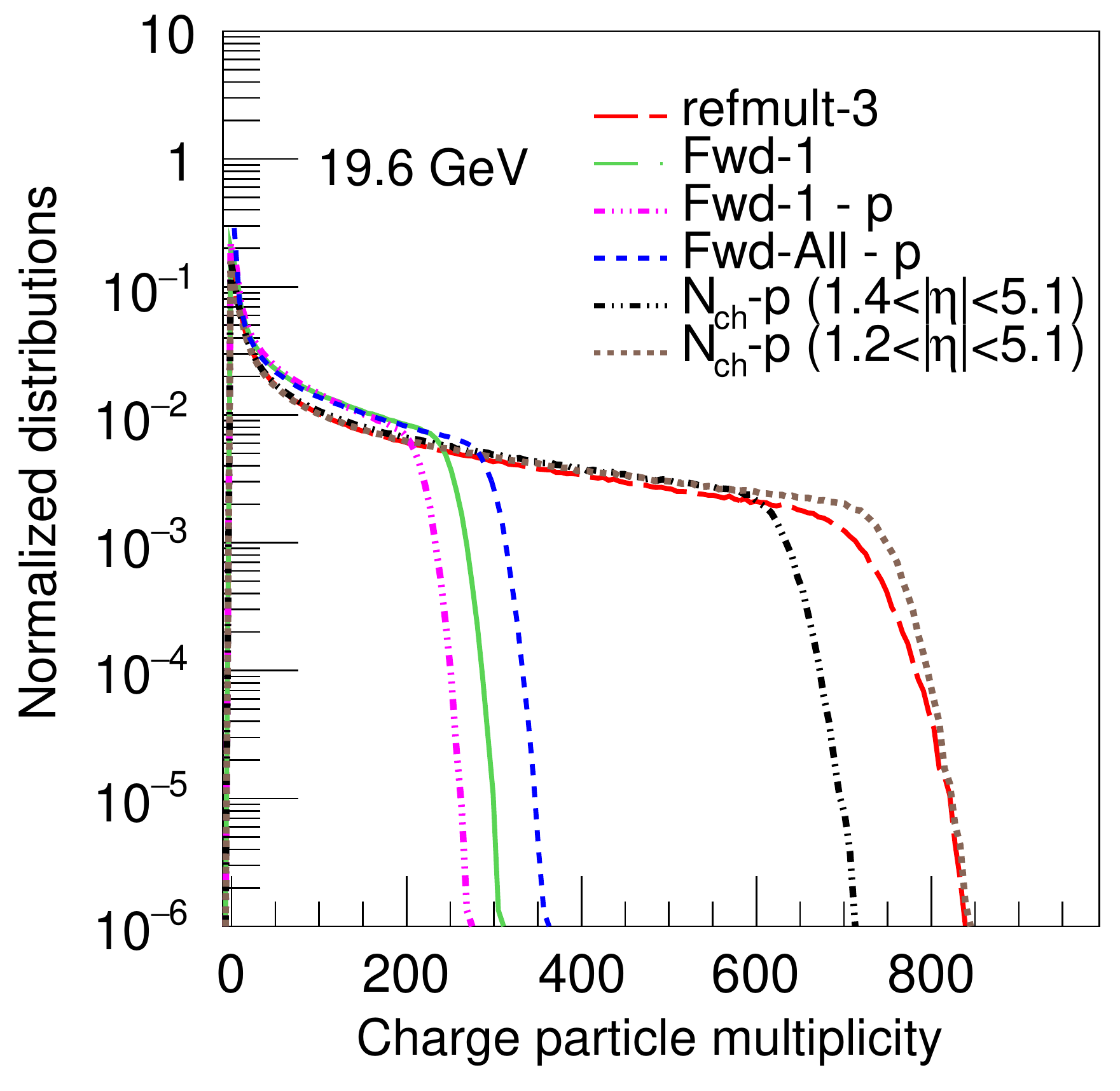}
	\caption{(Color online) Normalized distributions for charged particle multiplicities in different $\eta$-window in Au+Au collisions at \sNN = 19.6 from UrQMD model.}
	\label{mul19}
\end{figure}

\begin{figure*}[htp!]
	\centering 
	\includegraphics[width=0.9\textwidth]{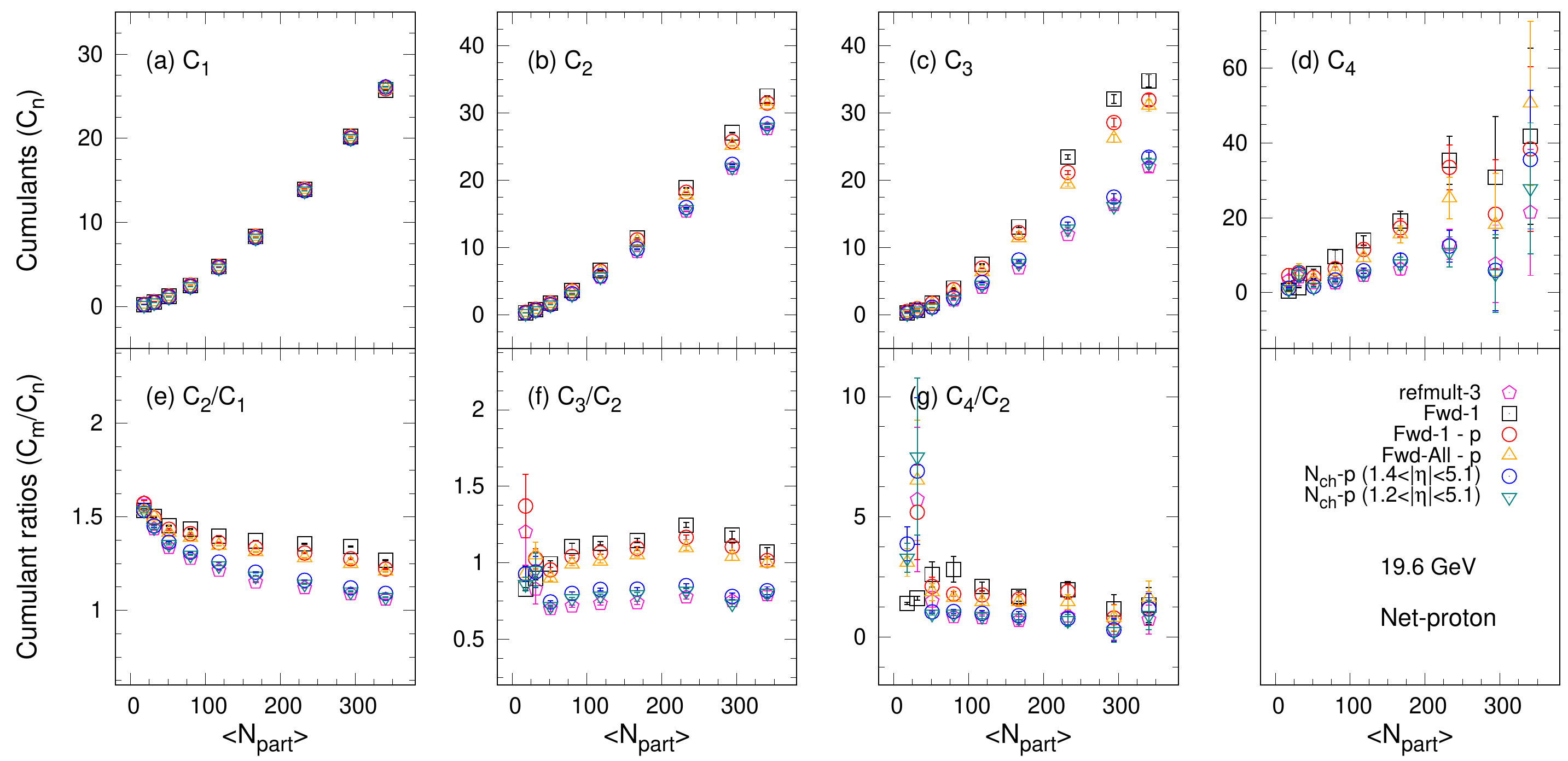}
	\caption{(Color online) Centrality dependence of cumulants ($C_{1} \sim C_{4}$) and cumulant ratios ($C_{2}/C_{1}$, $C_{3}/C_{2}$ and $C_{4}/C_{2}$) of net-proton multiplicity distributions within $|y|<0.5$ and $0.4<p_{T}<2.0$ GeV/$c$ in Au+Au collisions at \sNN= 19.6 GeV for different centrality selection methods. }
	\label{ratio19}
\end{figure*}

\begin{figure}[htp!]
	\centering 
	\includegraphics[width=0.4\textwidth]{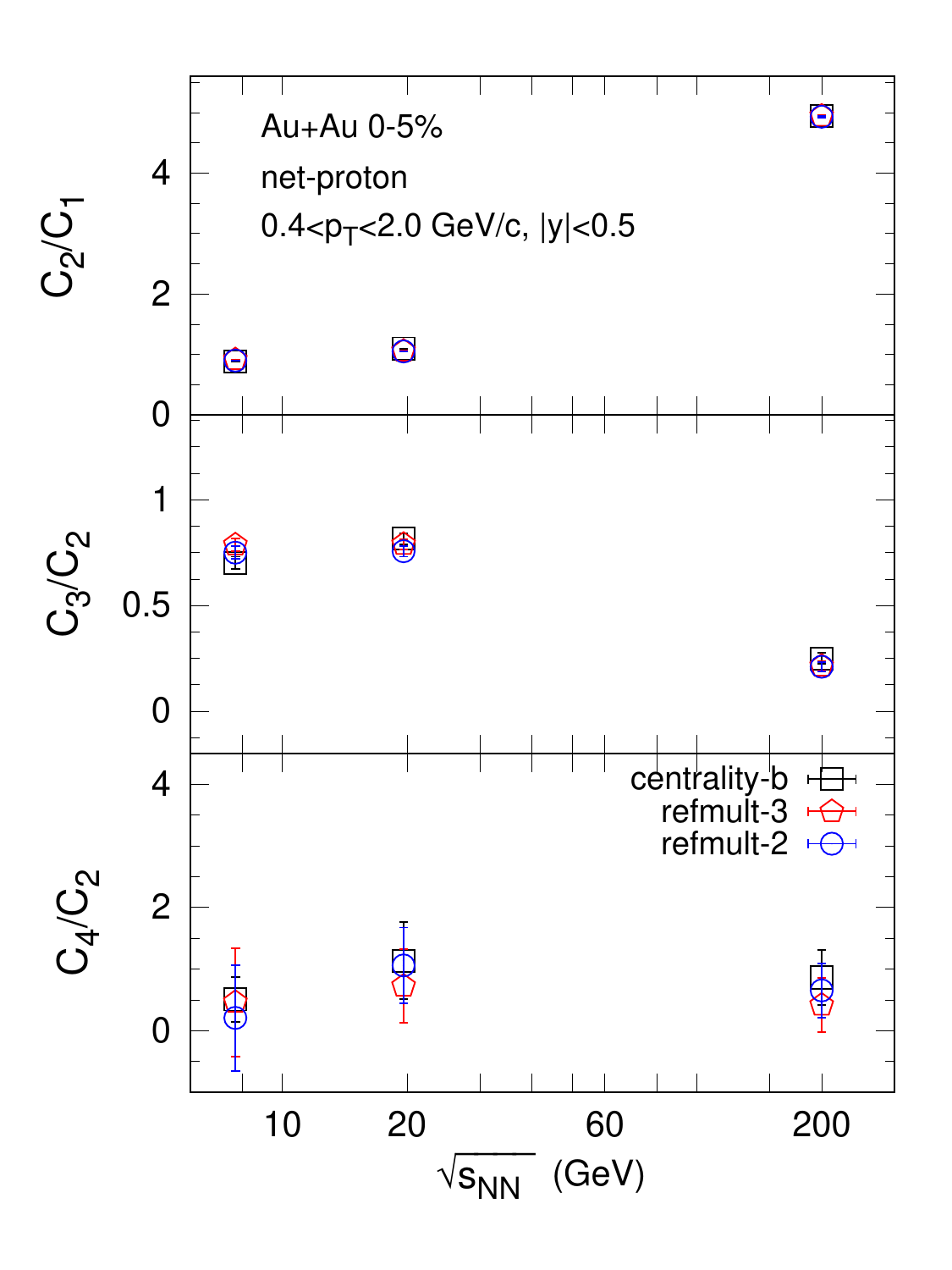}
	\caption{(Color online) Collision energy dependence of the cumulant ratios ($C_{2}/C_{1}$, $C_{3}/C_{2}$ and $C_{4}/C_{2}$) of net-proton 
	multiplicity distributions within $|y|<0.5$ and $0.4<p_{T}<2.0$ GeV/$c$ in 0-5\% most central Au+Au collisions for different centrality selection methods.}
	\label{Cnratio_energy}
\end{figure}

Figure~\ref{cumulants} shows the centrality dependence of cumulants ($C_{1}$ to $C_{4}$) of net-proton multiplicity distributions in Au+Au collisions at \sNN = 7.7, 19.6 and 200 GeV. The collision centralities are represented by the average number of participant nucleons ($\langle N_{part} \rangle$). We use a Monte Carlo Glauber model~\cite{Abelev:2008ab,Miller:2007ri} to estimate $N_{\rm part}$ similar to conventional cumulant analysis ~\cite{Aggarwal:2010wy,Adamczyk:2013dal,Luo:2015doi,Adamczyk:2014fia,Adamczyk:2017wsl,Adam:2019xmk}. 
The statistical uncertainties are obtained using analytical error propagation method~\cite{kendall1963advanced, Luo:2011tp,Luo:2014rea}. The statistical uncertainties mainly depends on the variance of the respective distributions and the number of events.  All the cumulants show a linear dependence as a function of $\langle N_{part} \rangle$. However $C_{1}$ and $C_{2}$ at \sNN = 7.7 GeV show an opposite trend for "Fwd-All" centrality case. This is because at 7.7 GeV within $2.1 < |\eta| < 5.1$ most of the charged particles are spectator protons. So in this region, larger $N_{ch}$ percentile corresponds to peripheral collisions not central collisions. However, if we substract protons from "Fwd-All" centrality definitions then the $C_{1}$ matches to other cases. We also observed that the cumulants values based on "refmult-2" and "refmult-3" centrality definitions are consistent for all three energies. For lower beam energies (\sNN = 7.7 and 19.6 GeV) higher order cumulants ($C_{3}$ and $C_{4}$) from "Fwd-All" and "Fwd-All-p"centrality definition are deviated from the cumulants using "refmult-2" and "refmult-3" centrality definition. This is because of the poor centrality resolution in "Fwd-All" and "Fwd-All-p"centrality definitions due to smaller multiplicity distribution and/or spectator contribution. For \sNN = 19.6 GeV,  we found the values of higher order cumulants from "refmult-1" are smaller than the results from "Fwd-All-p" centrality definition. This is caused by the autocorrelation effect in "refmult-1" centrality, as the centrality resolution of "Fwd-All-p" is better than the case of "refmult-1". Meanwhile, at 19.6 GeV, we found the higher order cumulants from "refmult-2" and "refmult-3" centrality cases are smaller than the results from "Fwd-All-p". We will discuss more later in this chapter (Fig. \ref{mul19}-\ref{Cnratio_energy}) showing that this is due to better centrality resolution in the refmult-2/refmult-3 than the "Fwd-All-p" case and not caused by the autocorrelation effects in refmult-2 and refmult-3 centrality definitions. 
At \sNN = 200 GeV, the cumulants from forward centrality definitions ("Fwd-All" and "Fwd-All-p") are consistent with the results from the centralities defined at central region ("refmult-2" and "refmult-3"). This comparison indicates that the autocorrelation effects in the centrality definitions of "refmult-2" and "refmult-3" are not significant within statistical uncertainties in Au+Au collisions at \sNN = 200 GeV within UrQMD model calculations.

Figure~\ref{ratio} shows the $\langle N_{part} \rangle$ dependence of cumulant ratios ((a) $C_{2}/C_{1} = \sigma^{2}/M$, (b) $C_{3}/C_{2} = S\sigma$ and (c) $C_{4}/C_{2} = \kappa\sigma^{2}$) of net-proton multiplicity distributions in Au+Au collisions at three different collision energies. At \sNN = 200 GeV, all the cumulants ratios in different centrality selection sets are consistent with each other. As we go towards lower energies, the effect of centrality selection start to play important role and cumulant ratios are deviating from each other.  The results in centrality definition sets from forward region show more deviates due to the poor centrality resolution caused by spectator proton contributions. 

As shown in Fig.~\ref{cumulants} and ~\ref{ratio}, the higher order cumulants and cumulant ratios for "refmult-2" and "refmult-3" centrality cases are smaller than 
the results from forward region centrality definition "Fwd-All-p".  We argue this is due to the better centrality resolution of "refmult-2" and "refmult-3" centrality definitions than the case of "Fwd-All-p". In Fig.~\ref{mul19}, we show the charged particle multiplicity distributions in various $\eta$-window in Au+Au collisions at \sNN = 19.6 from UrQMD model. We found that the charged particle multiplicity from "refmult-3" centrality is much larger than the forward centrality definition "Fwd-All-p".  This will cause larger volume fluctuations with "Fwd-All-p" centrality definition than the "refmult-3" case. To justify this argument, two new centralities were defined for Au+Au collisions at \sNN = 19.6 GeV from UrQMD with wider $\eta$ range in the forward region, which are charged particle multiplicities (excluding protons) within $1.4<|\eta|<5.1$ and $1.2<|\eta|<5.1$. 
By doing this, the charged particle multiplicities in the two new centrality definitions are much larger than the "Fwd-All-p" case and are similar as the multiplicities used in the "refmult-3" centrality definition. In Fig.\ref{ratio19}, the higher order cumulant and cumulant ratios from the two new centralities are very close to each other with the results from the "refmult-3" case and are much smaller than the "Fwd-All-p" centrality definition. It supports the argument that the large discrepancy between the results based on the "refmult-3" from central region and the "Fwd-All-p" from forward region are originated from the poor centrality resolution of "Fwd-All-p" definition. One may notice that there are small differences observed in $C_{2}/C_{1}$ and $C_{3}/C_{2}$ at non-central Au+Au collisions between "refmult-3" and "$N_{ch}-p$ ($1.2<|\eta|<5.1$)" cases, which might be due to some remaining autocorrelation in "refmult-3" case.
The comparison between the results of "Fwd-1" and "Fwd-1-p" implies that the spectator protons have a wide $\eta$ distribution in the forward region and can distort the centrality resolution even with a small fraction.

Figure~\ref{Cnratio_energy} shows the energy dependence of the cumulant ratios ($C_{2}/C_{1}$, $C_{3}/C_{2}$ and $C_{4}/C_{2}$) of net-proton multiplicity distributions in Au+Au collisions at \sNN= 7.7, 19.6 and 200 GeV for three centrality definition methods. The "centrality-b" is used to represent the centrality definition by using impact parameter. 
The centrality bin width correction has been applied to suppress volume fluctuations within wide centrality bin as discussed in section III. For "centrality-b", we first calculated cumulants in each 0.1 fm bin and then weight the cumulants by the number of events in each 0.1 fm bin over a desired centrality class as discussed in equation~\ref{cbwc}. 
Based on the UrQMD model study, we found that the results in 0-5\% most central Au+Au collisions from the "refmult-3" and "refmult-2" centralities are consistent with the results from "centrality-b" definition, which is directly related to the initial collision geometry. 
This comparison further confirms that the "refmult-3" and "refmult-2" centrality definitions are robust to be used for studying the net-proton fluctuations in heavy-ion collisions. 

\section{Summary}
The cumulants of net-proton multiplicity distributions are important observables to probe the signature of the QCD critical point in heavy-ion collisions. In this work, we studied the centrality dependence cumulants (up to fourth order) and the cumulants ratios ($C_{2}/C_{1}$, $C_{3}/C_{2}$ and $C_{4}/C_{2}$) of net-proton multiplicity distributions at \sNN = 7.7, 19.6 and 200 GeV for different centrality selection methods. These centralities were defined by the charged particle multiplicities from different central or forward region. We found that the mixture of the spectator protons and produced charged particles can distort the centrality definition with charged particle multiplicities at forward region and worsen the centrality resolution. Particularly, the situation will be worse at lower energies, where most of the spectator protons will be detected by STAR EPD. In the simulation, we demonstrated in detail that the centrality resolution can be significantly improved by excluding the spectator protons from the charged particle multiplicity used for centrality definition at forward region. However, in the STAR experiment, the EPD doesn't have particle identification capability, it would be very challenge to isolate these spectator protons from the produced charged particle and improve the centrality resolution. On the other hand, we found the higher order cumulants calculated from "refmult-3" and "refmult-2" centralities are consistent with the results 
from centralities defined by charged particles multiplicity (excluding protons) at forward region and the "centrality-b". It means the suspected autoautocorrelationcorrelation effects in the centrality definition of "refmult-3" and "refmult-2" at mid-rapidity are not significant within the statistical uncertainties within UrQMD model calculations. 
For EPD centrality definition, one could use more differential information ($\eta$-segmented multiplicity within the EPD detector) and deep learning technique~\cite{yuri_qm2019}. Our work will serve as a baseline for the centrality selection of the fluctuation analysis in future relativistic heavy-ion collision experiment.

\section{Acknowledgement}
 We thank ShinIchi Esumi, Jiangyong Jia, Mike Lisa, Bedangadas Mohanty, Tapan Kumar Nayak, Toshihiro Nonaka, and Nu Xu for stimulating discussion. This work is supported by the National Key Research and Development Program of China (2018YFE0205201),  the National Natural Science Foundation of China (No.11828501, 11575069, 11890711 and 11861131009).  NRS is supported by the Fundamental Research Funds of Shandong University.

\bibliography{centralityStudy}

\end{document}